\documentstyle[aps,multicol,epsf]{revtex}
\def\d{{\rm d}}
\begin{document}
\title{Path Integral Approach to the Dynamic Casimir Effect with
Fluctuating Boundaries}
\author{Ramin Golestanian}
\address{Institute for Advanced Studies in Basic Sciences,
Zanjan   45195-159, Iran}
\author{Mehran Kardar}
\address{Department of Physics, Massachusetts Institute of
Technology, Cambridge, MA 02139}
\date{\today}
\maketitle
\begin{abstract}
A path integral formulation is developed for the {\it dynamic}
Casimir effect. It allows us to study small deformations in
{\it space and time} of the perfectly reflecting (conducting)
boundaries of a cavity. The mechanical response of the
intervening vacuum is calculated to linear order in the
frequency--wavevector plane, using which a plethora of interesting
phenomena can be studied. For a single corrugated
plate we find a correction to mass at low frequencies,
and an effective shear viscosity at high frequencies
that are both anisotropic. The anisotropy is set by the wavevector
of the corrugation. For two plates, the mass
renormalization is modified by a function of the ratio between the
separation of the plates and the wave-length of corrugations.
The dissipation rate is not modified for frequencies below
the lowest optical mode of the cavity, and there is a resonant
dissipation for {\it all frequencies} greater than that.
In this regime, a divergence in the response function implies
that such high frequency deformation modes of the cavity can not be
excited by any macroscopic external forces. This phenomenon is
intimately related to resonant particle creation. For particular
examples of two corrugated
plates that are stationary, or moving uniformly in the lateral
directions,
Josephson-like effects are observed. For capillary waves on the surface
of mercury a renormalization to surface tension, and sound velocity is
obtained.

\end{abstract}
%%
%\begin{multicols}{2}%
%%
\section{Introduction and Summary}\label{sIntro}

The Casimir effect \cite{Casimir,Mostepa,Milloni,Krech,RMP} is a
surprising
phenomenon in which quantum fluctuations of the 
electromagnetic field in the vacuum, 
subject to the boundary conditions imposed by two conducting plates
of area $A$ and separation $H$, lead to an attractive force
between them, given by
\begin{equation} \label{casimirforce}
F_{\rm static}(H)=-\frac{\pi^{2}}{240}\times \hbar c \,\frac{A}{H^{4}}.
\end{equation}
This force, which provides a direct link between quantum field
theory and the macroscopic world, has been measured experimentally 
(using a torsion pendulum) between a gold plate, and a gold plated
sphere,
to the accuracy of within \%5 \cite{Lamor}. 
Related fluctuation-induced phenomena occur in various areas of physics
ranging from 
cosmology to statistical mechanics of phase transitions
\cite{Mostepa,Krech}. 

Since the pioneering work of Casimir, various modifications to the
standard
boundary conditions have been studied.
One generalization, relevant to experimental measurements of the force, 
is to introduce deformations (roughness) of the surfaces. 
For example, in Ref.\cite{BalDup} a multiple scattering approach is used
to 
compute the interactions for arbitrary geometry in a perturbation series
in the curvature. 
In another phenomenological approach, introduced in Ref.\cite{Bordag},
small 
deviations from plane parallel geometry are treated by
using an additive summation of van der Waals--type pair-wise potentials. 
The former method is difficult to implement, while the latter is not
always reliable
due to the additivity assumption \cite{RMP}.

New phenomena emerge for moving boundaries in the generalization to the 
{\it dynamic Casimir effect} 
\cite{Moore,Fulling,Jaekel,Neto,Neto2,Cavity,Meplan,Lambrecht,Davis,Ford}.
The creation of photons by moving mirrors was first obtained
by Moore\cite{Moore} for a 1 dimensional cavity.
Fulling and Davis\cite{Fulling} found a corresponding  force
proportional to the third time derivative of the mirror displacement.
Their methods specifically make use of the conformal symmetries
of the 1+1 dimensional space time, and hence are not applicable for
the more realistic case of 1+3 dimensions \cite{causality}.

Another approach to the dynamic Casimir effect starts with the 
fluctuations in the radiation pressure on a plate\cite{Neto,Neto2}.
The fluctuation--dissipation theorem is then used to obtain the 
mechanical response function, whose 
imaginary part gives the dissipation in response to motion. 
This method does not suffer from the causality problems of earlier 
approaches \cite{causality}, and can be applied to all dimensions. 
For example, the force in 1+3 dimensional space-time depends on 
the fifth time derivative of the displacement.
Several other approaches have focused on the emission of photons 
by a vibrating cavity \cite{Cavity,Meplan,Lambrecht}.
This is generally too small to be experimentally detectable
\cite{Davis,Knight}; the most promising set-up is the resonant 
production of photons when the mirrors vibrate at the optical 
resonance frequency of the cavity\cite{Davis}.
More recently, the radiation due to vacuum fluctuations of a
collapsing bubble has been proposed as a possible explanation
for the intriguing phenomenon of sonoluminescense
\cite{Eberlein,Knight}.
A good review of the topic including more extensive references
can be found in Ref.\cite{Barton}.

A number of authors have further discussed to notion of {\it frictional
forces}:
Using conformal methods in 1+1 dimensions, Ref. \cite{Dodonov1} finds
a friction term
\begin{equation} \label{frictionforce}
F_{\rm friction}(H)= \alpha \;F_{\rm static}(H)\;\left({\dot{H} \over
c}\right)^{2},
\end{equation}
for slowly moving boundaries, where $\alpha$ is a numerical constant 
that only depends on dimensionality.
The additional factor of $(v/c)^2$ would make detection of
this force yet more difficult.
There are a few attempts to calculate forces (in higher dimensions) 
for walls that move {\it laterally}, i.e. parallel to each other
\cite{Levitov,Mkrt,Pendry}: 
It is found that boundaries that are not ideal conductors, experience a
friction 
as if the plates are moving in a viscous fluid. 
The friction has a complicated dependence
on the frequency dependent resistivity of the plates, and vanishes
in the cases of ideal (nondissipating) conductors or dielectrics.
The ``dissipation'' mechanism for this ``friction'' is by inducing eddy
currents 
in the nonideal conductors, and thus distinct from the Casimir effect.

In this paper we present a path integral formulation, applicable to all
dimensions,  for the  problem of perfectly reflecting
mirrors that undergo small dynamic deformations.
Although the original
Casimir problem with parallel plates can be tackled with such a path
integral
approach (see Appendix \ref{Acasimir}), it is much easier to solve 
by the standard operator method, followed by expansions in modes
appropriate to the specified geometry and boundary conditions.
However, if one wishes to study the effect of deformations of
either static (fixed roughness) or dynamic nature, the standard
method is hardly tractable \cite{Bordag,Dodonov2}. In
contrast, the path integral method is much better suited
to handling
deformations. The only limitation of the method is that it
provides the result as a perturbative expansion in powers of
the ratio between the deformation field and the average
separation of the plates. In the present context, a great advantage
of this method is that
it does not make any distinction between ``time''
and ``space'' coordinates provided that appropriate Wick rotations
are performed. Consequently, one can consider time dependent
boundary conditions with no additional complications.

Using this procedure, we calculate the mechanical response function in
the
frequency--wavevector domain.
It is defined as the ratio between the induced force
and the deformation field, in the linear regime.
 From the response function we extract a plethora of interesting
results, some of which we list here for the specific example
of lateral vibrations of uniaxially corrugated plates:
\begin{itemize}
\item{\bf (1)} A single plate with corrugations of wavenumber ${\bf k}$,
vibrating at frequencies $\omega\ll ck$, obtains  {\it anisotropic}
corrections to its mass. Thus the effective mass of a plate depends on
its shape!
\item{\bf (2)} For $\omega\gg ck$, there is dissipation
due to a frequency dependent anisotropic shear viscosity,
i.e. a type of `friction' in the vacuum.
\item{\bf (3)} A second plate at a separation $H$ modifies
the mass renormalization by a function of $kH$, but does not change
the dissipation for frequencies $\omega^2<(ck)^2+(\pi c/H)^2$.
\item{\bf (4)} For all frequencies higher than this first optical normal
mode
of the cavity, the mechanical response is infinite, implying 
that such modes can not be excited by any finite external force.
This is intimately connected to the resonant particle creation 
reported in the literature\cite{Cavity,Davis}. 
It is also an example of radiation from neutral bodies, a purely quantum
effect.
\item{\bf (5)} A phase angle $\theta$ between two similarly 
corrugated plates results in Josephson--like effects: 
A static force proportional to $\sin (\theta)$, and an oscillating force
for a uniform relative velocity.
\item{\bf (6)} There is a (minute) correction to the velocity
of capillary waves on the surface of mercury due to a small
change in its surface tension.
\end{itemize}

The rest of the paper is organized as follows. In Sec.\ref{sDynamic}
we develop the formalism to calculate the effective action for
time dependent small deformation fields.
In Sec.\ref{sKernels} the behavior of the mechanical response
functions are examined in the frequency--wavevector domain.
We discuss the important special case of lateral
oscillations of rough plates in Sec.\ref{sSpecial},
and Sec.\ref{sConclus} gives the conclusions of the paper,
and suggests avenues for future explorations. In Appendix \ref{Acasimir}
the path integral method is applied to the standard Casimir effect
due to the electromagnetic field, while some details of the 
calculations are included in Appendix \ref{Adetail}.

\section{Path integral formulation of the dynamic Casimir effect}
\label{sDynamic}

In this section we introduce a quantization method for a field, subject
to boundary conditions on surfaces which undergo dynamic
shape fluctuations.
Although one can tackle the gauge invariant electromagnetic
vector field by this formalism (see Appendix \ref{Acasimir}),
we use for simplicity the ``scalar electrodynamics'' model
used most commonly in the literature \cite{Moore,Dodonov2}.
We consider a scalar field described by the classical action ($c=1$)
\begin{equation}
S=\frac{1}{2} \int \d^d X \;\partial_{\mu}\phi(X)
                        \partial_{\mu}\phi(X),\label{action}
\end{equation}
where summation over $\mu=1,\cdots,d$ in a $d$ dimensional
space-time is understood, and we
have made a Wick rotation by introducing the imaginary time
variable $X^d=i t$. We want to quantize the field $\phi$
subject to the constraint that it vanishes on a set
of manifolds (objects) embedded in space-time. The manifolds are
described by  functions
$X^{\mu}=X^{\mu}_{\alpha}(y_{\alpha})$;
a $D_{\alpha}$-dimensional manifold embedded in $d$-dimensional space
is parametrized by 
$y_{\alpha}\equiv (y^1_{\alpha},\cdots,y^{D_{\alpha}}_{\alpha})$.
Following
Ref.\cite{LiK}, the partition function of the system is written as
\begin{eqnarray}
{\cal Z}&=&\frac{1}{{\cal Z}_0}\int {\cal D}\phi(X)
                \prod_{\alpha=1}^{n} \prod_{y_{\alpha}}
                \delta\left(\phi\left(X_{\alpha}(y_{\alpha})
                \right)\right)\;\exp\left\{-\frac{1}{\hbar}S[\phi]
                \right\},\label{Z1}  \\
&=&\frac{1}{{\cal Z}_0}\int {\cal D}\phi(X)\prod_{\alpha=1}^{n}
{\cal D}\psi_{\alpha}(y_{\alpha})\; \exp\left\{
        -\frac{1}{\hbar}S[\phi] +i \sum_{\alpha}\int \d y_{\alpha}
        \sqrt{g_{\alpha}}\;\psi_{\alpha}(y_{\alpha})
        \phi\left(X_{\alpha}(y_{\alpha})\right)\right\},\nonumber
\end{eqnarray}
where ${\cal Z}_0$ is the partition function for the system
with no boundaries, and $g_{\alpha}$ is the determinant of
the induced metric on the $\alpha$'th manifold \cite{Diff},
\begin{equation}
g_{\alpha,ij}=\frac{\partial X^{\mu}_{\alpha}}{\partial y_{\alpha}^i}
              \frac{\partial X^{\mu}_{\alpha}}{\partial y_{\alpha}^j}
              ,\label{induced}
\end{equation}
required to make the integration measure over the manifolds
reparametrization invariant.

The integration over the field $\phi$ can be performed,
and the resulting expression for the partition function is
\begin{equation}
{\cal Z}=\int \prod_{\alpha} {\cal D}\psi_{\alpha}(y_{\alpha})
        \exp\left\{-S_1[\psi_{\alpha}(y_{\alpha})]\right\},
        \label{Z2}
\end{equation}
where
\begin{equation}
S_1[{\bf \psi}]\equiv {\bf \psi}^T M {\bf \psi}=\sum_{\alpha,\beta}
  \int \d y_{\alpha}\sqrt{g_{\alpha}} \d y_{\beta}\sqrt{g_{\beta}}
  \psi_{\alpha}(y_{\alpha}) G^d(X_{\alpha}(y_{\alpha})
  -X_{\beta}(y_{\beta})) \psi_{\beta}(y_{\beta}),\label{Mdef}
\end{equation}
and $G^d(X-X')=(-\partial_{\mu}\partial_{\mu})^{-1}_{XX'}$.
Integrating over the Gaussian fields $\{\psi_{\alpha}\}$
gives the final result of
\begin{equation}
\ln {\cal Z}=-\frac{1}{2} \ln \det \{M(X_{\alpha}(y_{\alpha}))\}.
        \label{lnZ}
\end{equation}

We next focus on the specific example of two $D$ (space-time)
dimensional surfaces in $d=D+1$ space-time with average separation
$H$, and small deformations. The surfaces are thus
parametrized by $X_1({\bf x})=({\bf x},h_1({\bf x}))$ and
$X_2({\bf x})=({\bf x},H+h_2({\bf x}))$, where $h_{1}({\bf x})$ and
$h_{2}({\bf x})$ are the deformations. One can read off the
matrix $M$ from Eq.(\ref{Mdef}) as
\begin{equation}
M({\bf x},{\bf y})=\left[
        \begin{array}{cc}
        \sqrt{g_1({\bf x})} \sqrt{g_1({\bf y})}
        \;G^d({\bf x}-{\bf y},h_1({\bf x})-h_1({\bf y}))
        & \sqrt{g_2({\bf x})} \sqrt{g_1({\bf y})}
        \;G^d({\bf x}-{\bf y},H+h_2({\bf x})-h_1({\bf y})) \\
        \sqrt{g_1({\bf x})} \sqrt{g_2({\bf y})}
        \;G^d({\bf x}-{\bf y},H+h_2({\bf y})-h_1({\bf x}))
        & \sqrt{g_2({\bf x})} \sqrt{g_2({\bf y})}
        \;G^d({\bf x}-{\bf y},h_2({\bf x})-h_2({\bf y}))
        \end{array} \right],\label{M}
\end{equation}
in which $\sqrt{g_{\alpha}({\bf x})}=\sqrt{1+(\nabla h_{\alpha})^2}$
where the gradient is with respect to the plane (space-time) coordinates
(see Eq.(\ref{induced})). Following Ref.\cite{LiK}, we perform a
perturbative expansion of the matrix in terms of the deformation
fields, as
\begin{equation}
M({\bf x},{\bf y})=M_0({\bf x},{\bf y})+\delta M({\bf x},{\bf y}).
        \label{Mpert}
\end{equation}
The matrix
\begin{equation}
M_0({\bf x},{\bf y})=
\left[ \begin{array}{cc}
        G^d({\bf x}-{\bf y},0)& G^d({\bf x}-{\bf y},H) \\
        G^d({\bf x}-{\bf y},H)& G^d({\bf x}-{\bf y},0)
        \end{array} \right],\label{M0}
\end{equation}
describes two flat plates and is only a
function of the differences of the coordinates. Hence it can be
transformed using a Fourier basis into
\begin{equation}
M_0({\bf p},{\bf q})=
\left[ \begin{array}{cc}
        G^d({\bf p})& G^d({\bf p},H) \\
        G^d({\bf p},H)& G^d({\bf p})
        \end{array} \right] \;(2\pi)^D \delta^D({\bf p}+{\bf q})
        ,\label{M0p}
\end{equation}
where
\begin{eqnarray}
G^d({\bf p})&=&\int \d^D {\bf x} \;G^d({\bf x},0) \;
        {\rm e}^{i{\bf p}\cdot{\bf x}},\label{Gp} \\
G^d({\bf p},H)&=&\int \d^D {\bf x} \;G^d({\bf x},H) \;
        {\rm e}^{i{\bf p}\cdot{\bf x}},\nonumber
\end{eqnarray}
are the Fourier-transformed Green's functions.

The part of the partition function that depends on the deformation
fields
can be written as $\ln {\cal Z}_h=-\frac{1}{2} \ln \det
\{1+M_0^{-1} \delta M \}$. Expanding everything in powers of
$h({\bf x})$, and keeping only up to second order terms, we
obtain (for details of the calculations see Appendix \ref{Adetail})
\begin{eqnarray}
\ln {\cal Z}_h&=&\frac{1}{2}\int \d^D {\bf x} \d^D {\bf y}
\;K({\bf x}-{\bf y})\;\left[h_1({\bf x})h_1({\bf y})
+h_2({\bf x})h_2({\bf y}) \right]   \label{Zh1} \\
&-&\frac{1}{2}\int \d^D {\bf x} \d^D {\bf y}
\;Q({\bf x}-{\bf y})\;\left[h_1({\bf x})h_2({\bf y})
+h_1({\bf y})h_2({\bf x})\right].   \nonumber
\end{eqnarray}
The kernels in the above equation are defined as
\begin{eqnarray}
K({\bf x})&=& \partial_{z}^2 G({\bf x},0)
F_1({\bf x})+F_1({\bf x}) F_5({\bf x})
+F_4({\bf x}) F_6({\bf x}), \label{KQ} \\
Q({\bf x})&=& \partial_{z}^2 G({\bf x},H)
F_4({\bf x})+F_2({\bf x})^2+F_3({\bf x})^2 , \nonumber
\end{eqnarray}
where
\begin{eqnarray}
F_1({\bf x})&=&\int \frac{d^D {\bf p}}{(2\pi)^D}
\frac{G^d({\bf p})}{{\cal N}({\bf p})}
\; {\rm e}^{i{\bf p}\cdot{\bf x}}, \label{F1-F6} \\
F_2({\bf x})&=&\int \frac{d^D {\bf p}}{(2\pi)^D}
\frac{G^d({\bf p})}{{\cal N}({\bf p})}\;
\frac{\partial G^d({\bf p},H)}{\partial H}
\; {\rm e}^{i{\bf p}\cdot{\bf x}}, \nonumber \\
F_3({\bf x})&=&\int \frac{d^D {\bf p}}{(2\pi)^D}
\frac{G^d({\bf p},H)}{{\cal N}({\bf p})}\;
\frac{\partial G^d({\bf p},H)}{\partial H}
\; {\rm e}^{i{\bf p}\cdot{\bf x}}, \nonumber \\
F_4({\bf x})&=&\int \frac{d^D {\bf p}}{(2\pi)^D}
\frac{G^d({\bf p},H)}{{\cal N}({\bf p})}\;
{\rm e}^{i{\bf p}\cdot{\bf x}}, \nonumber \\
F_5({\bf x})&=&\int \frac{d^D {\bf p}}{(2\pi)^D}
\frac{G^d({\bf p})}{{\cal N}({\bf p})}\;
\left(\frac{\partial G^d({\bf p},H)}{\partial H}\right)^2
\; {\rm e}^{i{\bf p}\cdot{\bf x}}, \nonumber \\
F_6({\bf x})&=&\int \frac{d^D {\bf p}}{(2\pi)^D}
\frac{G^d({\bf p},H)}{{\cal N}({\bf p})}\;
\left(\frac{\partial G^d({\bf p},H)}{\partial H}\right)^2
\; {\rm e}^{i{\bf p}\cdot{\bf x}}, \nonumber
\end{eqnarray}
and
\begin{equation}
{\cal N}({\bf p})=\left[G^d({\bf p})\right]^2-
                \left[G^d({\bf p},H)\right]^2. \label{Np}
\end{equation}

We now focus on the case of interest in the Casimir problem which
corresponds to $d=4$ and $D=3$. The Green's functions can be
calculated from the definition following Eq.(\ref{Mdef}), and
using Eqs.(\ref{Gp}) as
\begin{eqnarray}
G^4({\bf p})&=&\frac{1}{2p} ,\label{G4} \\
G^4({\bf p},H)&=&\frac{1}{2p}\;{\rm e}^{-pH} ,\nonumber \\
\partial_{z}^2 G^4({\bf x},0)&=&-\frac{1}{2\pi^2}\;
\frac{1}{x^4} ,\nonumber \\
\partial_{z}^2 G^4({\bf x},H)&=&-\frac{1}{2\pi^2}
\;\frac{x^2-3 H^2}{(x^2+H^2)^3}.\nonumber
\end{eqnarray}

To obtain the behavior due to the time dependence of shape
fluctuations of the surfaces, we rotate back to ``real''
time variables, changing Eq.(\ref{Zh1}) to
\begin{eqnarray}
\ln {\cal Z}_h=&-&\frac{1}{2}\int \d t \d t' \d^2 x \d^2 y
\;K(x-y,it-it')\;\left[h_1(x,t)h_1(y,t')
+h_2(x,t)h_2(y,t') \right]   \label{Zh2} \\
&+&\frac{1}{2}\int \d t \d t' \d^2 x \d^2 y
\;Q(x-y,it-it')\;\left[h_1(x,t)h_2(y,t')
+h_1(y,t')h_2(x,t)\right].   \nonumber
\end{eqnarray}
One can diagonalize the above expression using the Fourier
transformations
\begin{equation}
h_{\alpha}(x,t)=\int \frac{\d \omega \d^2 q}{(2\pi)^3}\;
{\rm e}^{-i \omega t+i q \cdot x}\;h_{\alpha}(q,\omega),\label{hqw}
\end{equation}
for the fields. The corresponding transformations for the kernels
result in
\begin{eqnarray}
{\cal K}(q,\omega)&=&\int \d t \d^2 x\;{\rm e}^{-i\omega t+iq\cdot x}
K(x,it) \label{Kqw} \\
&=&-i \int_{-\infty}^{\infty} \d \tau \int \d^2 x
\;{\rm e}^{-\omega \tau+iq\cdot x} 
K\left(\sqrt{x^2+\tau^2}\right) \nonumber \\
&\equiv&-i\;A_{+}(q,\omega), \nonumber
\end{eqnarray}
and
\begin{eqnarray}
{\cal Q}(q,\omega)&=&\int \d t \d^2 x\;{\rm e}^{-i\omega t+iq\cdot x}
Q(x,it) \label{Qqw} \\
&=&-i \int_{-\infty}^{\infty} \d \tau \int \d^2 x
\;{\rm e}^{-\omega \tau+iq\cdot x} 
Q\left(\sqrt{x^2+\tau^2}\right) \nonumber \\
&\equiv&-i\;A_{-}(q,\omega), \nonumber
\end{eqnarray}
where we have performed a rotation of the integration contours from
$i t$ to $\tau$. The symmetric appearance of the argument
$\sqrt{x^2+\tau^2}$
in Eqs.(\ref{Kqw}) and (\ref{Qqw}) is due to the underlying Lorentz
invariance
of the theory.
Finally, the resulting expression for the effective action, defined via
$\ln {\cal Z}=i S_{\rm eff}/\hbar$, reads
\begin{eqnarray}
S_{\rm eff}=\frac{\hbar c}{2} \int\frac{\d \omega \d^2 q}{(2\pi)^3}\;
\left\{A_{+}(q,\omega)\left(|h_1(q,\omega)|^2+|h_2(q,\omega)|^2\right)
\hskip 4 cm \right. \label{Seff} \\
\left. \hskip 2 cm -A_{-}(q,\omega)\left(h_1(q,\omega)
h_2(-q,-\omega)+h_1(-q,-\omega)h_2(q,\omega)\right) \right\}.\nonumber
\end{eqnarray}

\section{Analytic structure of the response kernels} \label{sKernels}

The kernels $A_{\pm}(q,\omega)$ are closely related to the mechanical
response of the system. By substituting the expressions of Eq.(\ref{G4})
in Eqs.(\ref{KQ}-\ref{Np}) and (\ref{Kqw},\ref{Qqw}), 
after some manipulations, one obtains
\begin{equation}
A_{\pm}(q,\omega)=\frac{\pi^2}{64 H^5} \int\limits_{-\infty}^{+\infty}
        \d s \left(\frac{\sin\left[2 \sqrt{q^2-\omega^2/c^2} H s/\pi
\right]}
        {2 \sqrt{q^2-\omega^2/c^2} H s/\pi}\right) g_{\pm}(s),
        \;\;\; \mbox{for\ $\omega < c q$}.  \label{A+-}
\end{equation}
For $\omega > c q$, the result is obtained by {\it analytic
continuation}
of the above. The need for such analytic continuation is a distinction
between
Minkowski and Euclidean spaces, and did not occur in Ref.\cite{LiK}.
In Eq.(\ref{A+-}),
\begin{equation}
g_{\pm}(s)=\left(\frac{1}{s^3} \pm \frac{\cosh s}{\sinh^3 s}\right)^2
+\frac{\sinh^2 s}{\cosh^6 s} \pm 2 s \frac{\sinh s}{\cosh^3 s}
\left[\frac{s^2-3\pi^2/4}{(s^2+\pi^2/4)^2}\right],\label{g+-}
\end{equation}
are two functions of the dimensionless parameter $s$, and
upper and lower cut-offs for the space-time are understood in
case of divergences.
It can be seen that the kernels are functions of
the separation $H$, but depend on ${\bf q}$ and $\omega$ only
through the combination $Q^2=q^2-\omega^2/c^2$.
Rather than elaborating on the closed forms of the kernels,
we shall describe their behavior in various regions of the
parameter space.

The kernels can be calculated exactly for a single mirror.
It is simpler to take the
$H \rightarrow \infty$ limit in the original Eqs.(\ref{KQ}-\ref{G4}),
and (\ref{Kqw},\ref{Qqw}). 
Additional manipulations lead to $A_{-}^{\infty}(q,\omega)=0$, and
\begin{equation}
A_{+}^{\infty}(q,\omega)=\left\{
        \begin{array} {cl}
        - \frac{1}{360 \pi^2 c^5}(c^2 q^2-\omega^2)^{5/2}
        & \mbox{for $\omega < c q$} , \\ 
         \\
        i\frac{{\rm sgn}(\omega)}{360 \pi^2 c^5} (\omega^2- c^2
q^2)^{5/2}
        & \mbox{for $\omega > c q$},
        \end{array} \right. \label{A+infty}
\end{equation}
where ${\rm sgn}(\omega)$ is the sign function.
While the effective action and the corresponding response are real
for $Q^2>0$, they become purely imaginary for $Q^2<0$.
The analytic continuation to this regime is not unique. 
However, if we require  causality for the response, 
we are led to the choice in which  the imaginary part is 
an odd function, leading to the inclusion of the sign  function
in the above formula.
An imaginary response function signals dissipation of energy
\cite{Neto}, presumably
by creation of photons as shown in the next section \cite{Lambrecht}.
For the special case of $q=0$, Eq.(\ref{A+infty}) agrees with the
results
obtained previously\cite{Neto} for flat
mirrors. (Note that Ref.\cite{Neto} considers the electromagnetic field
rather than a scalar field, accounting
for the discrepancy in the numerical prefactor. We have also done
the same calculations in 1+1 dimensions and have reproduced the
result of Refs.\cite{Fulling,Jaekel}, who also use scalar fields,
with the correct coefficient.)

In the presence of a second plate (i.e. for finite $H$),
the situation is more complicated. For $Q^2>0$, we can examine the
behavior of the integrand in Eq.(\ref{A+-}) and find
that the kernels are {\it real} and {\it finite} in this regime.
However, for
$Q^2<0$ the situation is different. Applying contour integration,
we can transform the integral in Eq.(\ref{A+-}) to the form
\begin{equation}
{\cal I}= 2 \pi i \sum {\rm residues} -{\cal CI},\label{Res}
\end{equation}
where ${\cal CI}$ is the contribution from the semicircle at
infinity. The residues can be easily shown to be finite and
real, and add to up to a finite and well behaved sum.
Examining the contributions from the semicircle at infinity, we
find terms with both real and imaginary parts, that generically
behave as
${\cal CI} \sim \exp[(K-2)L/H]/[K(L/H)^3]$, with $K=2 Q' H/\pi$,
$Q'=i Q$.  $L$ is the radius of the semicircle, presumably
corresponding to an upper cut-off in space-time.
Now we can see that in the limit $L \rightarrow \infty$, the
situation drastically changes at $K=2$. While ${\cal CI}$ vanishes
for $K \leq 2$, it diverges with both real and imaginary parts
for $K >2 $, where the inequality is understood in its strict
sense.

To summarize, we can subdivide the parameter space of the kernels
into three different regions, as depicted in Fig.~1.
In region I ($Q^2>0$ for any $H$),
the kernels are finite and real, and hence there is no dissipation.
In region IIa where $-\pi^2/H^2 \leq Q^2 <0$, the
$H$-independent part of $A_{+}$ is imaginary, while the
$H$-dependent parts of both kernels are real and finite.
(This is also the case at the boundary $Q^2=-\pi^2/H^2$.)
The dissipation in this regime is simply the sum of what would
have been observed if the individual plates were decoupled,
and unrelated to the separation $H$.
By contrast, in region IIb where $Q^2 < -\pi^2/H^2$, both
kernels diverge with infinite real and imaginary parts.
This $H$-dependent divergence extends all the way to the negative
$Q^2$ axis, where it is switched off by a $1/H^5$ prefactor.
Note that some care is necessary in the order of limits for
$(L,H)\to\infty$.

Divergence of the kernels might be argued to signal a breakdown
of the perturbation theory. However, one should note that the
exponential divergence of the kernels is an essential singularity 
that can not be ``cured'' by standard renormalization techniques.
It is in fact an artifact of the unphysical assumption of perfect
reflectivity for the mirrors at {\it all} frequencies. 
Any natural frequency cut-off for the reflectivity of the mirror will
round off this divergence \cite{Lambrecht}, leading to a finite
mechanical response. In other words, the imperfection in reflectivity
at low frequencies provides a leakage mechanism for the cavity 
that stops the resonant energy build-up.
Exponential creation of photons (with observation time) for 
perfectly reflecting boundaries has also been reported in the 
literature for the case of a one dimensional cavity \cite{Meplan}, 
using a completely different approach.

\section{Corrugated Plates} \label{sSpecial}
We now concentrate on a concrete example, and examine the lateral
vibrations of surfaces with fixed (time independent) roughness,
such as two corrugated plates.
We assume that the first plate undergoes a lateral
motion described by ${\bf r}(t)$, while the second plate is
stationary. The deformations of the plates are thus described by
$h_1({\bf x},t)=h_1({\bf x}-{\bf r}(t))$ and
$h_2({\bf x},t)=h_2({\bf x})$.
The lateral force exerted on the first plate is obtained from
$f_i(t)=\delta S_{\rm eff}/\delta r_i(t)$.
Using Eq.(\ref{Seff}), and within linear response,
this is given by
\begin{equation}
f_i(\omega)=
\chi_{ij}(\omega)\;r_j(\omega)+f_i^{0}(\omega),\label{rough1}
\end{equation}
where the ``mechanical response tensor'' is
%\end{multicols}
\begin{equation}
\chi_{ij}(\omega)=\hbar c \int \frac{\d^2 q}{(2\pi)^2}
\;q_i q_j \left\{\left[A_{+}(q,\omega)-A_{+}(q,0)\right]\;|h_1({\bf
q})|^2
+\frac{1}{2} A_{-}(q,0) \left(h_1({\bf q})h_2(-{\bf q})+h_1(-{\bf q})
h_2({\bf q})\right) \right\}, \label{rough}
\end{equation}
and there is a residual (static) force
\begin{equation}
f_i^{0}(\omega)=-\frac{\hbar c}{2}\;2\pi\delta(\omega)\;
\int \frac{\d^2 q}{(2\pi)^2}
\;i q_i A_{-}(q,0) \left(h_1({\bf q})h_2(-{\bf q})
-h_1(-{\bf q})h_2({\bf q})\right). \label{fi0w}
\end{equation}
%\begin{multicols}{2}
We can calculate the dissipation rate
\begin{eqnarray}
P&=&\lim_{T \rightarrow \infty}\frac{1}{T} \int\limits_{-T/2}^{T/2}
\d t \;\dot{\bf r}(t) \cdot {\bf f}(t),\\ \label{pow1}
&=&\frac{1}{T} \int\limits_{-\infty}^{+\infty} \frac{\d \omega}{2\pi}
\; i\omega\; \chi_{ij}(\omega)\;r_i(-\omega)r_j(\omega),\nonumber \\
&=&-\frac{1}{T} \int\limits_{-\infty}^{+\infty} \frac{\d \omega}{2\pi}
\; \omega\; {\rm Im}\chi_{ij}(\omega)\;r_i(-\omega)r_j(\omega),\nonumber
\end{eqnarray}
in which we have used the fact that real (imaginary) part of a
response function is an even (odd) function of frequency.
Using the above formulas, we can predict various interesting effects
for corrugated plates, that appear due to quantum fluctuations of
vacuum.

\subsection{Single plate phenomena} \label{sSingle}

For a single corrugated plate, whose deformation is defined as
$h({\bf x})=d \cos{{\bf k}\cdot {\bf x}}$, one can easily calculate
the response tensor using the explicit formulas in Eq.(\ref{A+infty}).
In the low-frequency limit, i.e. when $\omega\ll c k$, we can expand
the result in powers of $\omega$. This gives
$\chi_{ij}=\delta m_{ij}\omega^2+O(\omega^4)$, where
$\delta m_{ij}=A\hbar d^2 k^3 k_i  k_j/(288\pi^2 c)$,
and can be regarded as corrections to the mass of the plate
\cite{Newton}.
(Cut-off dependent mass corrections also appear, as in Ref.
\cite{Barton}.)
Note that these mass corrections are {\it anisotropic}, with
\begin{eqnarray} \label{singledm}
\delta m_{\parallel}&=&A\hbar  k^5 d^2/(288\pi^2 c),\\
\delta m_{\perp}&=&0. \nonumber
\end{eqnarray}
Parallel and perpendicular components are
defined with respect to ${\bf k}$, and $A$ denotes the area of the
plates.
The mass correction is proportional to $\hbar$, and hence inherently
very small. For example, if we have a macroscopic
sample with $d\approx \lambda=2\pi/k \approx 1$mm, density
$\approx 15 {\rm gr/cm}^3$, and thickness $t\approx 1$ mm, we
find $\delta m/m \sim 10^{-34}$. Even for deformations of a microscopic
sample of atomic dimensions (close to the limits of the applicability
of our continuum representations of the boundaries),  $\delta m/m$
can only be reduced to around $10^{-10}$. While the actual
changes in mass are immeasurably small, the hope is that its
{\it anisotropy} may be more accessible, say by comparing
oscillation frequencies of a corrugated plate in two orthogonal
directions.

For $\omega\gg ck$, which is the high-frequency limit, the response
function is imaginary, and we can define a frequency dependent
effective shear viscosity by
$\chi_{ij}(\omega)=-i\omega\eta_{ij}(\omega)$. This viscosity is also
anisotropic, with
\begin{eqnarray} \label{singleeta}
\eta_{\parallel}(\omega)&=&\hbar A k^2 d^2 \omega^4/(720\pi^2 c^4),\\
\eta_{\perp}(\omega)&=&0. \nonumber
\end{eqnarray}
Note that in 1+3 dimensions the dissipation is proportional to the
fifth time derivative of displacement, and there is no dissipation
for a uniformly accelerating plate. However, a freely oscillating
plate will undergo a damping of its motion. The characteristic decay
time for a plate of mass $M$ is $\tau\approx 2M/\eta$. For the
macroscopic plate of the previous paragraph, vibrating at a frequency
of $\omega\approx 2 c k$ (in the $10^{12}$Hz range), the decay time is
enormous, $\tau\sim 10^{18}$s. However, since the decay time
scales as the fifth power of the dimension, it can be reduced to
$10^{-12}$s, for plates of order of 10 atoms. However, the required
frequencies in this case (in the $10^{18}$Hz range) are very large.
Also note that for the linearized forms to remain valid in this high
frequency regime, we must require very small amplitudes,
so that the typical velocities involved $v\sim r_0\omega$, are
smaller than the speed of light. These difficulties can be somewhat
overcome by considering resonant dissipation in the presence of
a second plate.

\subsection{Double plate phenomena} \label{sDouble}

With two plates at an average distance $H$, the results are
qualitatively the same for frequencies less than the natural
resonance of the resulting cavity. There is a renormalization of mass
in region I, and dissipation appears in region IIa, of Fig.~1.
However, the mass renormalization at low frequencies ($\omega\ll ck$)
is now a function of both $k$ and $H$, with a crossover from the
single plate behavior for $k H \sim 1$. In the limit of $k H \ll 1$,
we obtain
\begin{eqnarray} \label{doubledm}
\delta m_{\parallel}&=&\hbar A B k^2 d^2/48 c H^3 ,\\
\delta m_{\perp}&=&0, \nonumber
\end{eqnarray}
with
\begin{equation} \label{B}
B=\int\limits_{0}^{\infty} \; \d s \; s^2
\;\left[-\frac{4}{s^6}+g_{+}(s)\right]
\approx -0.452448.
\end{equation}
Compared to the single plate,
there is an enhancement by a factor of $(kH)^{-3}$ in
$\delta m_\parallel$. The effective dissipation in region IIa is simply
the sum of those due to individual plates, and contains no $H$
dependence.

There are additional interesting phenomena resulting from resonances.
We find that both real and imaginary parts of $A_{\pm}(q,\omega)$,
diverge for $\omega^2/c^2 > q^2+\pi^2/H^2$, that corresponds to
region IIb of Fig.~1.
In the example of corrugated plates, we replace $q$ by $k$ to obtain
a continuous spectrum of frequencies with diverging dissipation.
(Note that the imaginary part of the response function gives the
dissipation.)
Related effects have been reported in the literature for 1+1
dimensions\cite{Cavity,Meplan,Lambrecht,Davis}, but occuring at
a {\it discrete} set of frequencies $\omega_n=n \pi c/H$ with
integer $n\geq2$. These resonances occur when the frequency of the
external perturbation matches the natural normal modes of the cavity,
thus exciting quanta of such modes. In one space dimension,
these modes are characterized by a discrete set of wavevectors
that are integer multiples of $\pi/H$. The restriction to $n\geq2$ is a
consequence of quantum  electrodynamics being a `free' theory
(quadratic action): only two-photon states can be excited subject to
conservation of energy. Thus the sum of the frequencies of the two
photons should add up to the external frequency\cite{Lambrecht}.

In higher dimensions, the appropriate parameter is the combination
$\omega^2/c^2-q^2$. From the perspective of the excited photons,
conservation of momentum requires that their two momenta add up
to $q$, while energy conservation restricts the sum of their frequencies
to $\omega$. The in-plane momentum $q$, introduces a continuous
degree of freedom: the resonance condition can now be satisfied for a
continuous spectrum, in analogy with optical resonators.
In Ref.\cite{Lambrecht}, the lowest resonance frequency is found to
be $2\pi c/H$ which seems to contradict our prediction. However,
the absence of $\omega_1=\pi c/H$ in 1+1 D is due to a vanishing
prefactor\cite{Lambrecht}, which is also present in our calculations.
However, in exploring the continuous frequency spectrum in higher
dimensions, this single point is easily bypassed, and there is a
divergence for all frequencies satisfying
$\omega^2/c^2 > q^2+\pi^2/H^2$, where the inequality holds in its
strict sense.

Resonant dissipation has profound consequences for motion of
plates. It implies that due to quantum fluctuations of vacuum,
{\it components of motion with frequencies in the range of
divergences cannot be generated by any finite external force}!
The imaginary parts of the kernels are proportional to the total
number of excited photons \cite{Lambrecht}. Exciting these degrees
of motion must be accompanied by the generation of an infinite
number of photons; requiring an infinite amount of  energy, and thus
impossible. However, as pointed out in Ref.\cite{Lambrecht}, the
divergence is rounded off by assuming finite reflectivity and
transmissivity for the mirrors. Hence, in practice, the restriction is
softened and controlled by the degree of ideality of the mirrors in
the frequency region of interest.

\subsection{Josephson-like effects} \label{sJosephson}

Consider two plates corrugated at the same wavelength, and
separated at a distance $H$, whose deformations are described by
$h_1({\bf x},t)=d_1 \; \cos[{\bf k} \cdot ({\bf x}-{\bf r}(t))]$ and
$h_2({\bf x},t)=d_2 \; \cos[{\bf k} \cdot {\bf x}]$.
For the case of uniform motion of the plate, namely,
${\bf r}(t)= {\bf v}\;t+ {\bf r}_0$, we can find the
full expression for the lateral force  as
\begin{equation}
{\bf f}(t)=\frac{\delta S_{\rm eff}}{\delta {\bf r}(t)}
=\frac{\hbar c A}{2} A_{-}(k,0) {\bf k} d_1 d_2
        \sin[{\bf k} \cdot {\bf r}(t)].\label{uniform}
\end{equation}
While in previous expressions we performed a expansion to linear
order in ${\bf r}(t)$, in Eq.(\ref{uniform}) the complete expression is
calculated using the time translational invaraince of a uniformly
moving object.
For two stationary plates (${\bf v}=0$) with a phase mismatch of
$\alpha={\bf k} \cdot {\bf r}_0$, there is a (time independent)
lateral force
\begin{equation}
{\bf F}_{dc}=\frac{\hbar c A}{2} A_{-}(k,0) {\bf k} d_1 d_2
        \sin{\alpha},
\label{JosephsonDC}
\end{equation}
which tends to keep the plates 180 degrees out of phase, i.e.
mirror symmetric with respect to their mid-plane.
This should be regarded as a correction to the static Casimir
force due to the deformations of the plates.
The dependence on the sine of the phase mismatch is
reminiscent of the DC Josephson current in superconductor
junctions, the force playing a role analogous to the current.
(The amplitudes $d_1$ and $d_2$ are similar to the wave
functions in SIS junctions.)
Note that we could have equally well obtained this result using
the constant term in Eq.(\ref{fi0w}).

For uniform motion, there is an analog for the AC Josephson
effect, with velocity (the variable conjugate to force) playing
the role of voltage. Setting ${\bf r}_0=0$, we find
\begin{equation}
{\bf F}_{ac}=\frac{\hbar c A}{2} A_{-}(k,0) {\bf k} d_1 d_2
        \sin[({\bf k} \cdot {\bf v})\; t],
\label{JosephsonAC}
\end{equation}
where $A_{-}(k,0)=1/H^{5}\; g(k H) $, with $g$ a (dimensionless)
slowly varying function of its argument. Note that the force
oscillates at a frequency $\omega={\bf k} \cdot {\bf v}$.
Actually both effects are a consequence of the attractive
nature of the Casimir force. It would be difficult to separate them
from similar forces resulting from say, van der Waals attractions.
One difference with (non-retarded) van der Waals force is in the
power-law fall-off. 
Another potential difference with additive attractive interactions
may be in the angular dependence of a static force. 
Such a difference in angular dependence due to the
non-additive nature of thermal  fluctuation-induced forces is 
discussed in Refs.\cite{GGK,RMP} for two rods on a membrane.
It is also likely to occur if we examine two rod-shaped
deformations on the two plates in the Casimir geometry.

\subsection{Capillary waves} \label{sCapillary}

In addition to solid corrugated plates, we can examine
deformations of fluid surfaces.
For example, consider the capillary waves on the surface
of mercury, with a conducting plate placed at a separation $H$
above the surface. The low frequency--wavevector expansion of the
kernel due to quantum fluctuations in the intervening vacuum starts
with quadratic forms $q^2$ and $\omega^2$. These terms result in
corrections to the (surface) mass density by
$\delta \rho=\hbar B/48 c H^3$, and to the surface
tension by $\delta \sigma=\hbar c B/48 H^3$ ($B$ is the numerical
constant calculated in Eq.(\ref{B})). The latter correction
is larger by a factor of $(c /c_s)^2$, and changes the
velocity $c_s$, of capillary waves by
$\delta c_s/c_s^{0}=\hbar c B/96 \sigma H^3$,
where $\sigma$ is the bare surface tension of mercury.
Taking $H\sim 1 $mm and $\sigma \sim 500$ dynes/cm, we find
another very small correction of $\delta c_s/c_s^{0} \sim 10^{-19}$.

\section{Conclusions and outlook} \label{sConclus}

In conclusion, we have developed a path integral formulation for the
study of quantum fluctuations in a cavity with dynamically deforming
boundaries. As opposed to previous emphasis on spectra of emitted
radiation, we focus on the mechanical response of the vacuum from which
we extract a variety of interesting mechanical effects.
Most of the predicted  dynamic Casimir phenomena, while
quite intriguing theoretically, appear to be beyond the reach of current
experiment: the most promising candidates are the anisotropy in mass,
and resonant dissipation. A major difficulty is  the necessity of
exciting deformation modes of macroscopic objects in the $10^{12}$Hz
or higher frequency range. However, as the static Casimir effect seems
more amenable to current experiments \cite{Lamor}, it would be
interesting to test the non-additive nature of this force
for plates with rod-shaped deformations.

The path integral method is quite versatile: 
It is better suited to treating deformed boundaries 
than  other techniques, such as that of Ref.\cite{Ford}. 
Moreover, unlike other approaches \cite{NetoPrivate}, 
it provides the dispersive part of the response, 
as well as its dissipative part.   
Furthermore, we can envision a number of future extensions
based on this formalism, some of which are listed below:
\begin{enumerate}
\item{It is worthwhile to repeat the calculations for electromagnetic gauge 
fields. 
While we expect the above results to remain qualitatively unchanged, 
the vector nature of EM field, as well its gauge symmetry, may lead to
novel additional phenomena.
The transversality of photons in vacuum, as well differences between transverse 
electric (TE) and transverse magnetic (TM) modes, are two examples
of the difference with scalar field, and may be important when considering 
the spectrum of  emitted radiation (see below).}

\item{There have been some efforts to calculate the spectrum of radiation 
emitted from a single fluctuating mirror, or by a cavity with fluctuating walls
\cite{Lambrecht,Knight}.
Our formalism allows calculations in the more realistic case of 3+1 
dimensional  undulating walls. 
(Presumably, it is simpler to excite surface deformations 
of the boundaries at high frequencies, rather than rigidly oscillating 
the wall as a whole.)
Radiation spectra can be obtained from two-point correlation functions
which may be calculated by extending the methods of this paper. 
Hopefully, the angular distribution of the spectrum will provide signatures 
that can help in the experimental detection of such weak radiation.}

\item{
New issues can be probed by extending the calculation to the next order in
the deformation fields.
For example, new dissipative terms may arise at the second order, 
resulting in a `viscosity' proportional to the  velocity-squared
as in Eq.(\ref{frictionforce}).}

\item{As in Ref.\cite{LiK}, it is possible to examine other geometries such as
fluctuating 1 and 2 dimensional manifolds, corresponding respectively to 
world-lines of a point particle and a wire in space-time.
We can then use our techniques to calculate the dynamic Casimir forces 
involved for fluctuating wires, and mutual interactions between moving particles,
wires,  and surfaces.}

\item{Dielectric superlattices are artificial media with nontrivial optical 
properties\cite{JDJ}.
Because of the dielectric contrast in photonic crystals, they are
excellent candidates for Casimir effects.
In particular, dimensional arguments suggest that a photonic crystal with lattice
spacing $\lambda$, will receive a density enhancement (presumably anisotropic)
of the order of  $\hbar /(c\lambda^4)$. 
While this is at best many orders of magnitude smaller than a typical density, 
it is none-the-less interesting to calculate the associated corrections to the 
mass tensor.
It is more promising to examine the coupling of the vibrations of such 
superlattices
to their optimal properties.
In particular, do mechanical deformations (phonons) modify the optical function 
of  a photonic crystal, due to quantum fluctuations of the vacuum?}

\item{An important generalization is to replace the assumption of perfectly
metallic walls with boundary conditions that describe  more realistic dielectric 
functions.}
\end{enumerate}

\acknowledgements
RG acknowledges many helpful discussions with M.R.H. Khajehpour,
B. Mashhoon, S. Randjbar-Daemi, and Y. Sobouti, and support from
the Institute for Advanced Studies in Basic Sciences, Gava Zang,
Zanjan, Iran.  MK is supported by the NSF grant DMR-93-03667.

\appendix

\section{Path integral method for the electromagnetic field}   
\label{Acasimir}

In this Appendix we introduce a path integral method for
the standard static Casimir effect due to quantum fluctuations of the
electromagnetic (EM) field. We start with the
action
\begin{equation}
S_0=-\frac{1}{4}\int \d^4 X \sqrt{-g} \;g^{\mu \alpha}
g^{\nu \beta} F_{\mu \nu} F_{\alpha \beta},\label{S0}
\end{equation}
where $F_{\mu \nu}=\partial_{\mu} A_{\nu}-\partial_{\nu}
A_{\mu}$, and $A_{\mu}=(A_0,-{\bf A})$ is the vector potential. We have
introduced the constant (space-time independent) metric as a source, to
enable
calculation of the
integral over the stress-energy tensor \cite{LanLif}, as
\begin{equation}
\left.\frac{\partial S}{\partial g^{\mu \nu}}\right|_{\eta^{\mu \nu}}
=\frac{1}{2} \int \d^4 X \;T_{\mu \nu}(X),\label{Tmunu}
\end{equation}
and $\eta^{\mu \nu}={\rm diag}(-1,1,1,1)$ is the flat space-time
metric.

We then take two perfectly conducting parallel plates
separated a distance $H$ along the $z$-direction. The boundary
condition subject to which we have to quantize the theory is
that the transverse electric field should vanish on the plates
\cite{Mostepa}. We should also treat the gauge freedom of
the action by adopting a proper gauge fixing procedure.
One can show that a possible proper quantization is achieved
by considering the following expression for the generating
function \cite{Chang}
\begin{equation}
Z(g)=\lim_{\beta \rightarrow 0} \frac{1}{Z_0} \int {\cal D} A(X)
\prod_{i,\alpha=1}^{2} \prod_{y_{\alpha}}\;
\delta \left(\partial_{0}A_i(X_{\alpha}(y_{\alpha}))-
\partial_{i}A_0(X_{\alpha}(y_{\alpha})) \right)\;
{\rm e}^{\frac{i}{\hbar} S_{\beta}[A]},\label{Zg}
\end{equation}
where
\begin{eqnarray}
S_{\beta} &\equiv &S_0[A]-\frac{\hbar}{2 \beta}
\int \d^4 X \sqrt{-g} \left(g^{\mu \nu} \partial_{\mu} A_{\nu}
\right)^2 \label{Sbeta} \\
&=&\int \d^4 X \sqrt{-g} \left[\frac{1}{2}A_{\mu}
\left(g^{\mu \nu} g^{\alpha \beta} \partial_{\alpha}
\partial_{\beta}-g^{\mu \alpha} g^{\nu \beta} \partial_{\alpha}
\partial_{\beta}(1-\frac{\hbar}{\beta}) \right) A_{\nu} \right].
\nonumber
\end{eqnarray}
Now one can easily calculate the average value for energy using
\begin{eqnarray}
-2 i \hbar
\left.\frac{\partial \ln Z(g)}{\partial g^{00}}\right|_{\eta^{\mu \nu}}
&=&\int \d^4 X \;\langle T_{00}(X) \rangle \label{<Tmunu>} \\
&=& A T \; {\cal E},\nonumber
\end{eqnarray}
where $A$ is the area of the plates, $T$ is the overall time
interval, and ${\cal E}$ is the Casimir energy.
We now represent the delta-functions using Lagrange multiplier fields,
as
\begin{equation}
Z(g)=\lim_{\beta \rightarrow 0} \frac{1}{Z_0} \int {\cal D} A(X)
\prod_{i,\alpha=1}^{2} {\cal D} \psi_{i,\alpha}(y_{\alpha})\;
\exp\left\{\frac{i}{\hbar} S_{\beta}[A]
+i \sum_{i,\alpha} \int \d y_{\alpha} \;\psi_{i,\alpha}(y_{\alpha})
\left(\partial_{0}A_i(X_{\alpha}(y_{\alpha}))-
\partial_{i}A_0(X_{\alpha}(y_{\alpha})) \right) \right\}.\label{Zg2} \\
\end{equation}
Performing the integration over the gauge field, yields
\begin{equation}
Z(g)=\lim_{\beta \rightarrow 0} \frac{1}{Z_0}
\int \prod_{i,\alpha=1}^{2} {\cal D} \psi_{i,\alpha}(y_{\alpha})\;
\exp\{-S[\psi_{i,\alpha}(y_{\alpha})]\},\label{Zg3}
\end{equation}
in which
\begin{eqnarray}
S[\psi]=\psi^{T} M \psi=\sum_{\alpha,\beta} \int \d y_{\alpha} \d
y_{\beta}
\;&&\left[\partial_{i} \psi_{i,\alpha}(y_{\alpha})
G_{00}(X_{\alpha}(y_{\alpha})-X_{\beta}(y_{\beta})) \partial_{j}
\psi_{j,\alpha}(y_{\alpha}) \right.
\label{Spsi} \\
&& + \partial_{0} \psi_{i,\alpha}(y_{\alpha})
G_{ij}(X_{\alpha}(y_{\alpha})
-X_{\beta}(y_{\beta})) \partial_{0} \psi_{j,\alpha}(y_{\alpha})
\nonumber \\
&& - \left. 2 \partial_{0} \psi_{i,\alpha}(y_{\alpha})
G_{0i}(X_{\alpha}(y_{\alpha})-X_{\beta}(y_{\beta})) \partial_{j}
\psi_{j,\alpha}(y_{\alpha}) \right],\nonumber
\end{eqnarray}
and the matrix Green's functions satisfy
\begin{equation}
-i\left[g^{\mu\nu}g^{\alpha\beta}\partial_{\alpha}\partial_{\beta}
-\left(1-{\hbar \over \beta}\right)
g^{\mu\alpha}g^{\nu\beta}\partial_{\alpha}\partial_{\beta} \right]
G_{\nu \lambda}(X-X')=\delta^{\mu}_{\lambda} \delta^{4}(X-X').
\label{MatGreen}
\end{equation}

The geometry of two parallel flat plates at a separation $H$ is
implemented
using the embeddings
$X_1 ({\bf x},t)=(t,{\bf x},0)$, and $X_2 ({\bf x},t)=(t,{\bf x},H)$.
Then we can read off the matrix $M$ from Eq.(\ref{Spsi}). In Fourier
space, it is diagonal and reads
\begin{equation}
M(p,q)=(2\pi)^3 \delta^{3}(p+q)
\;\left(p_i p_j M_{00}+p_0^2 M_{ij}+p_0 p_i M_{0j}
+p_0 p_j M_{0i}\right),\label{Mpqmat}
\end{equation}
in which
\begin{equation}
M_{\mu\nu}=
\left[ \begin{array}{cc}
        G_{\mu\nu}(p,0)& G_{\mu\nu}(p,H) \\
        G_{\mu\nu}(p,H)& G_{\mu\nu}(p,0)
        \end{array} \right],\label{Mmunu}
\end{equation}
and
\begin{eqnarray}
G_{\mu\nu}(p,H)&=&\int \d t \d^2 {\bf x} \;
G_{\mu\nu}(t,{\bf x},H) \;
        {\rm e}^{i{\bf p}\cdot{\bf x}-i p_0 t}.\label{Gmunup}
\end{eqnarray}
Having found the matrix $M$, we can calculate the generating
function from
\begin{equation}
\ln Z(g)=-\frac{1}{2} \ln \det \{M\}.\label{logdet}
\end{equation}
To this end, we only need to determine the matrix Green's
functions from Eq.(\ref{MatGreen}), which in momentum space
are given by
\begin{equation}
G_{\nu\lambda}(P)=\frac{-i}{g^{\alpha\beta} P_{\alpha}
P_{\beta}} \left[g_{\nu\lambda}+\left(\frac{\beta}{\hbar}-1\right)
\frac{P_{\nu} P_{\lambda}}{g^{\alpha\beta} P_{\alpha} P_{\beta}}\right],
\label{GnulambP}
\end{equation}
where $P_{\nu}$ is the four-momentum. At this point we can set
$\beta=0$,
which implements the (Lorentz) gauge choice without any difficulty.

To simplify our calculations, we take the specific form
$g^{\mu\nu}={\rm diag}(g^{00},1,1,1)$ for the metric. This is
sufficient for our purposes, as we are only need variations
with respect to $g^{00}$ to obtain the average energy.
Inserting the matrix Green's functions from
Eq.(\ref{GnulambP}), in Eqs.(\ref{Mpqmat}-\ref{Gmunup}), 
we obtain after some manipulations,
\begin{equation}
M(p,q)=(2\pi)^3 \delta^{3}(p+q)
\left[-i\left(\frac{p_i p_j}{g^{00}}+p_0^2 \delta_{ij}\right)\right]
\otimes
\left[\begin{array}{cc}
I(g^{00},p,0) & I(g^{00},p,H) \\
I(g^{00},p,H) & I(g^{00},p,0)
\end{array} \right],\label{MI}
\end{equation}
where
\begin{equation}
I(g^{00},p,H)= \frac{\exp[-H(g^{00} p_0^2+{\bf p}^2)^{1/2}]}
{2 (g^{00} p_0^2+{\bf p}^2)^{1/2}}.\label{Ieqn}
\end{equation}
We can next calculate the determinant to get
\begin{equation} \label{lastZg}
\ln Z(g)=- A T \int \frac{\d^3 p}{(2\pi)^3}\;
\ln\left(1-\exp\left[-2 H(g^{00} p_0^2+{\bf p}^2)^{1/2}\right]\right).
\end{equation}
Finally, we use Eq.(\ref{<Tmunu>}), and obtain
\begin{equation}
{\cal E}=-\frac{\pi^2}{720} \frac{\hbar c}{H^3},\label{ECasimir}
\end{equation}
which is the well known result for the Casimir energy \cite{Casimir}.

\section{Perturbation around flat plates} \label{Adetail}

In this Appendix we present the details of the calculations that
lead to Eq.(\ref{Zh1}).
We follow the steps and notations in the Appendix of Ref.\cite{LiK}.
We also generalize this work by including deformations of the second
plate,
and by using an invariant measure for integrations over the manifolds.
We also note that the final result of Eq.(\ref{Zh1}), corrects some
mistakes that appeared in Ref.\cite{LiK}.
To find $\delta M$ we expand the square roots, as well as the Green's
functions in Eq.(\ref{M}) up to second order in the deformation
\begin{eqnarray}
G^d({\bf x}-{\bf y}, h_{1,2}({\bf x})-h_{1,2}({\bf y}))
&=&G^d({\bf x}-{\bf y},0)+\frac{1}{2} \partial^2_{z}G^d({\bf x}-{\bf
y},0)
[h_{1,2}({\bf x})-h_{1,2}({\bf y})]^2,\label{Gd} \\
G^d({\bf x}-{\bf y}, H+h_2({\bf x})-h_1({\bf y}))
&=&G^d({\bf x}-{\bf y},H)+\partial_{z}G^d({\bf x}-{\bf y},H)
(h_2({\bf x})-h_1({\bf y})) \label{GdH} \\
&+&\frac{1}{2} \partial^2_{z}G^d({\bf x}-{\bf y},H)
[h_2({\bf x})-h_1({\bf y})]^2.\nonumber
\end{eqnarray}
After Fourier transformation one obtains (see Eq.(\ref{Mpert}))
\begin{eqnarray}
\delta M({\bf p},{\bf q})=
\left[ \begin{array}{cc}
        B_1({\bf p},{\bf q}) & A({\bf p},{\bf q}) \\
        A({\bf q},{\bf p}) & B_2({\bf p},{\bf q})
        \end{array} \right] ,\label{delM}
\end{eqnarray}
where
\begin{eqnarray}
A({\bf p},{\bf q})=\int \d^D {\bf x} \d^D {\bf y}
\;{\rm e}^{i{\bf p}\cdot{\bf x}+i{\bf q}\cdot{\bf y}}
\left\{\frac{1}{2}G^d({\bf x}-{\bf y},H) \left[(\nabla h_1({\bf x}))^2
+(\nabla h_2({\bf y}))^2 \right]  \right.  \label{ABpq} \\
+\partial_{z}G^d({\bf x}-{\bf y},H)
(h_2({\bf y})-h_1({\bf x})) \nonumber \\
\left. \hskip 3 cm +\frac{1}{2} \partial^2_{z}G^d({\bf x}-{\bf y},H)
[h_2({\bf y})-h_1({\bf x})]^2 \right\},\nonumber \\
B_{1,2}({\bf p},{\bf q})=\int \d^D {\bf x} \d^D {\bf y}
\;{\rm e}^{i{\bf p}\cdot{\bf x}+i{\bf q}\cdot{\bf y}}
\left\{\frac{1}{2}G^d({\bf x}-{\bf y},0) \left[(\nabla h_{1,2}({\bf
x}))^2
+(\nabla h_{1,2}({\bf y}))^2 \right]  \right. \nonumber \\
\left. \hskip 3 cm +\frac{1}{2} \partial^2_{z}G^d({\bf x}-{\bf y},0)
[h_{1,2}({\bf y})-h_{1,2}({\bf x})]^2 \right\}.\nonumber
\end{eqnarray}
The matrix $M_0$ is diagonal in Fourier space and can be
easily inverted as
\begin{eqnarray}
M_0^{-1}({\bf p},{\bf q})=\frac{1}{{\cal N}({\bf p})}
\left[ \begin{array}{cc}
        G^d({\bf p})& -G^d({\bf p},H) \\
        -G^d({\bf p},H)& G^d({\bf p})
        \end{array} \right] \;(2\pi)^D \delta^D({\bf p}+{\bf q})
        .\label{M0-1}
\end{eqnarray}
We then calculate $\ln {\cal Z}_h=-\frac{1}{2} \;{\rm tr} \ln(1+M_0^{-1}
\delta M)$, using the expansion of the logarithm up to the
second order. We also choose $\int \d^D {\bf x} \;h({\bf x})=0$
without loss of generality, leading to
\begin{eqnarray}
\ln {\cal Z}_h=-\frac{1}{2}\int \frac{\d^D {\bf p}}{(2\pi)^D}
\int \d^D {\bf x} \d^D {\bf y} \left\{\frac{G^d({\bf p})}{{\cal N}({\bf
p})}
\;{\rm e}^{i{\bf p}\cdot({\bf x}-{\bf y})} \left[G^d({\bf x}-{\bf y},0)
\left((\nabla h_1({\bf x}))^2+(\nabla h_2({\bf x}))^2 \right)
\right.\right. \label{bigZ} \\
\left.+\frac{1}{2} \partial^2_{z}G^d({\bf x}-{\bf y},0)
\left((h_1({\bf y})-h_1({\bf x}))^2+(h_2({\bf y})-h_2({\bf x}))^2
\right) \right] \nonumber \\
-\frac{G^d({\bf p},H)}{{\cal N}({\bf p})}
\;{\rm e}^{i{\bf p}\cdot({\bf x}-{\bf y})} \left[G^d({\bf x}-{\bf y},H)
\left((\nabla h_1({\bf x}))^2+(\nabla h_2({\bf x}))^2 \right) \right.
\nonumber \\
\left.\left. +\partial^2_{z}G^d({\bf x}-{\bf y},H)
\left(h_2({\bf y})-h_1({\bf x}) \right)^2 \right] \right\} \nonumber \\
+\frac{1}{2} \int \frac{\d^D {\bf p}}{(2\pi)^D}\frac{\d^D {\bf
q}}{(2\pi)^D}
\;\left(\frac{G^d({\bf p})G^d({\bf q})+G^d({\bf p},H)G^d({\bf q},H)}
{{\cal N}({\bf p}){\cal N}({\bf p})} \right) \hskip 4cm \nonumber \\
\times\; \left\{\int \d^D {\bf x} \d^D {\bf v}
\;{\rm e}^{i{\bf q}\cdot{\bf x}+i{\bf p}\cdot{\bf v}}
\; \partial_{z}G^d({\bf v}-{\bf x},H)h_2({\bf x})
-\int \d^D {\bf x} \d^D {\bf v}
\;{\rm e}^{i{\bf p}\cdot{\bf x}+i{\bf q}\cdot{\bf v}}
\; \partial_{z}G^d({\bf v}-{\bf x},H)h_1({\bf x}) \right\} \nonumber \\
\times\; \left\{\int \d^D {\bf y} \d^D {\bf u}
\;{\rm e}^{-i{\bf q}\cdot{\bf y}-i{\bf p}\cdot{\bf u}}
\; \partial_{z}G^d({\bf u}-{\bf y},H)h_2({\bf y})
-\int \d^D {\bf y} \d^D {\bf u}
\;{\rm e}^{-i{\bf p}\cdot{\bf y}-i{\bf q}\cdot{\bf u}}
\; \partial_{z}G^d({\bf u}-{\bf y},H)h_1({\bf y}) \right\}. \nonumber
\end{eqnarray}
Integration over ${\bf u}-{\bf y}$ and ${\bf v}-{\bf x}$ can now be
performed followed by integrations over ${\bf p}$ and ${\bf q}$,
resulting in
Eq.(\ref{Zh1}). Note that we have eliminated some
uninteresting terms that were proportional to $\int d^D{\bf p}$
and $\int d^D{\bf p} \;{\bf p}^2$. Such integrals are usually removed
in the framework of dimensional regularization \cite{Zinn}, and don't
play any important role. They appear since the ``constant'' infinite
energy that is usually neglected
when quantizing a field, now depends on the configuration of the
surfaces. When we dynamically deform the boundaries, this contribution
changes. These terms give rise to cut-off corrections to mass and
surface tension as discussed in Ref.\cite{Ford}

\begin{figure}
\epsfysize=3.0truein
\centerline{\epsffile{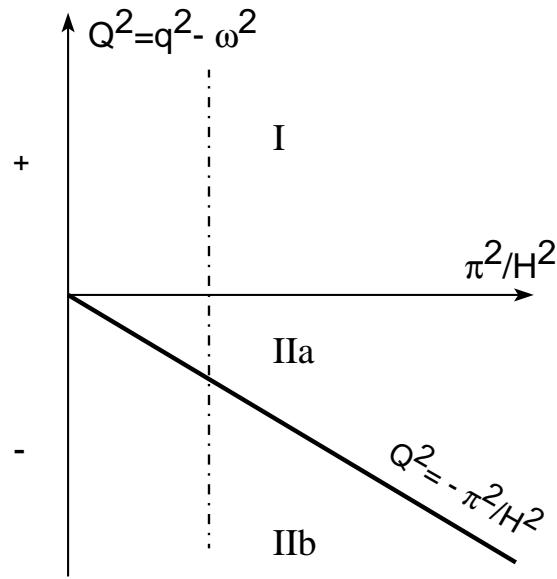}}
%Fig.1 Different regions of the $({\bf q},\omega)$ plane.
\caption{Different regions of the $({\bf q},\omega)$ plane.}
\label{Fig1}
\end{figure}

%\end{multicols}

%%
%\bibitem{CasX}
%See, e.g. J.N. Israelachvili and P.M. McGuigan, Science {\bf 241},
%6546 (1990).
%
%
%
%\bibitem{AM}
% N.W. Ashcroft and N.D. Mermin, {\it Solid State Physics}
% (Holt, Rinehart, and Winston, USA, 1976).
%%

\begin{references}

\bibitem{Casimir}
 H.B.G. Casimir, Proc. K. Ned. Akad. Wet. {\bf 51},
 793 (1948).

\bibitem{Mostepa}
 V.M. Mostepanenko and N.N. Trunov, Sov. Phys. Usp.
 {\bf 31}, 965 (1988).

\bibitem{Milloni}
 P.W. Milloni, {\it The Quantum Vacuum} (Academic Press, New York,
 1994).

\bibitem{Krech}
 M. Krech, {\it The Casimir effect in critical systems} (World
Scientific,
 Singapore, 1994).

\bibitem{RMP}
 M. Kardar and R. Golestanian, submitted to Rev. Mod. Phys.
(Colloquia)(1997).

\bibitem{Lamor}
 S.K. Lamoreaux, Phys. Rev. Lett. {\bf 78}, 5 (1997).

\bibitem{BalDup}
 R. Balian and B. Duplantier, Ann. Phys. (N.Y.){\bf 112}, 165 (1978).

\bibitem{LiK}
 H. Li and M. Kardar, Phys. Rev. Lett. {\bf 67}, 3275 (1991);
 Phys. Rev. A {\bf 46}, 6490 (1992).

\bibitem{Bordag}
 M. Bordag, G.L. Klimchitskaya, V.M. Mostepanenko, Phys. Lett. A
 {\bf 200}, 95 (1995).

\bibitem{Moore}
G.T. Moore, J. Math. Phys. {\bf 11}, 2679 (1970).

\bibitem{Fulling}
S.A. Fulling, and P.C.W. Davies, Proc. R. Soc. A {\bf 348}, 393
(1976).

\bibitem{Jaekel}
M.-T. Jaekel, and S. Reynaud, Phys. Lett. A {\bf 167}, 227 (1992).

\bibitem{Neto}
P.A. Maia Neto, and S. Reynaud, Phys. Rev. A {\bf 47}, 1639 (1993).

\bibitem{Neto2}
 P.A. Maia Neto and L.A.S. Machado, Braz. J. Phys. {\bf 25}, 324 (1995).
 
\bibitem{Cavity}
G. Calucci, J. Phys. A: Math. Gen. {\bf 25}, 3873 (1992);
C.K. Law, Phys. Rev. A {\bf 49}, 433 (1994);
V.V. Dodonov, Phys. Lett. A {\bf 207}, 126 (1995).

\bibitem{Meplan}
O. Meplan and C. Gignoux, Phys. Rev. Lett. {\bf 76}, 408
(1996).

\bibitem{Lambrecht}
A. Lambrecht, M.-T. Jaekel, and S. Reynaud, Phys. Rev. Lett.
{\bf 77}, 615 (1996).

\bibitem{Davis}
P. Davis, Nature {\bf 382}, 761 (1996).

\bibitem{Ford}
L.H. Ford and A. Vilenkin, Phys. Rev. D {\bf 25}, 2569 (1982).

\bibitem{causality}
The calculated force also has causality
problems reminiscent of the radiation reaction forces in classical
electron theory. However, it has been shown that
this problem, which is an artifact of the unphysical assumption of
perfect
reflectivity of the mirror at any frequency, can be resolved by
taking frequency dependent reflectivity and transmittivity functions
that respect the Kramers--Kronig relations\cite{Jaekel}.

\bibitem{Eberlein}
C. Eberlein, Phys. Rev. Lett. {\bf 76}, 3842 (1996); Phys. Rev. A
{\bf 53}, 2772 (1996).

\bibitem{Knight}
P. Knight, Nature {\bf 381}, 736 (1996).

\bibitem{Barton}
G. Barton and C. Eberlein, Ann. Phys. (N.Y.) {\bf 227}, 222 (1993).

\bibitem{Dodonov1}
 V.V. Dodonov, A.B. Klimov, and V.I. Man'ko, Phys. Lett. A
{\bf 142}, 511 (1989).

\bibitem{Levitov}
L.S. Levitov, Europhys. Lett. {\bf 8}, 499 (1989).

\bibitem{Mkrt}
 V.E. Mkrtchian, Phys. Lett. A {\bf 207}, 299 (1995).

\bibitem{Pendry}
 J.B. Pendry, preprint (1997) cond-mat/9707190.

\bibitem{Dodonov2}
 V.V. Dodonov, Phys. Lett. A {\bf 207}, 126 (1995).

\bibitem{Diff}
 See Appendix \ref{Adetail} for the differences between the present
formalism
 and that of Ref.\cite{LiK}.

\bibitem{Newton}
This identification of the effective mass is based simply on the naive
use of  a Newtonian equation of  motion for the plate.
This is criticized in Ref.\cite{Jaekel}, as the  equation of motion is 
in fact more complicated, involving also higher time derivatives.
However, our usage of the term is in line with other examples, 
such as the effective mass of an electron in a crystal.

\bibitem{GGK}
R. Golestanian, M. Goulian, and M. Kardar, Europhys. Lett. {\bf 33},
241 (1996); Phys. Rev. E {\bf 54}, 6725 (1996).

\bibitem{NetoPrivate}
 P.A. Maia Neto, private communication.

\bibitem{JDJ}
J. D. Joannopoulos, R. D. Meade, and J. N. Winn, {\it Photonic Crystals} 
(Princeton University Press, Princeton, 1995 ). 

\bibitem{LanLif}
 L.D. Landau and E.M. Lifshitz, {\it The classical theory
 of fields} (Pergamon Press, England, fourth edition, 1975).

\bibitem{Chang}
 S.J. Chang, {\it Introduction to quantum field theory}
 (World Scientific, Singapore, 1990).

\bibitem{Zinn}
 J. Zinn-Justin, {\it Quantum field theory and critical phenomena}
 (Oxford University Press, England, second edition, 1993).


\end{references}
\end{document}